\documentclass[english,aps,prl,twocolumn,oneside,floatfix]{revtex4}
\usepackage[T1]{fontenc}
\usepackage[latin9]{inputenc}
\usepackage{float}
\usepackage{graphicx}
\usepackage{amssymb}

\makeatletter

\usepackage{babel}
\makeatother

\begin{document}

\title{Quantum and Classical Chirps in an Anharmonic Oscillator}

\author{Yoni Shalibo$^{1}$, Ya'ara Rofe$^{1}$, Ido Barth$^{1}$, Lazar
Friedland$^{1}$, Radoslaw Bialczack$^{2}$, John M. Martinis$^{2}$
and Nadav Katz$^{1}$}

\affiliation{$^{1}$Racah Institute of Physics, The Hebrew University of Jerusalem,
Jerusalem 91904, Israel}

\affiliation{$^{2}$Department of Physics, University of California, Santa Barbara,
California 93106, USA}
\begin{abstract}
We measure the state dynamics of a tunable anharmonic quantum system,
the Josephson phase circuit, under the excitation of a frequency-chirped
drive. At small anharmonicity, the state evolves like a wavepacket
- a characteristic response in classical oscillators; in this regime
we report exponentially enhanced lifetimes of highly excited states,
held by the drive. At large anharmonicity, we observe sharp steps,
corresponding to the excitation of discrete energy levels. The continuous
transition between the two regimes is mapped by measuring the threshold
of these two effects.
\end{abstract}


\maketitle
Ever since the laws of quantum mechanics were formulated, there has
been an ongoing effort to explain the emergence of classical laws
in experimental systems. The first explanation by Bohr states that
these systems operate in the limit of large quantum numbers \cite{Bohr1920},
in which case they may be described by a wavepacket that on the average
follows the classical equations of motion \cite{Ehrenfest1927}. In
addition, coupling to uncontrolled, external degrees of freedom (decoherence),
is often related to the emergence of classicality \cite{Zurek2003}.
Recent experiments and calculations have demonstrated the quantum
to classical transition in oscillators, via noise saturation at
low temperature due to zero point fluctuations \cite{Katz2007,Murch2011},
and harmonic behavior at high temperatures in a cavity-QED system
\cite{Fink2010}.

In a classical anharmonic oscillator, such as a pendulum, the energy
expectation can be deterministically increased to large values if
the driving force is frequency-chirped and its amplitude is sufficiently
large. This phenomenon is commonly known as autoresonance \cite{Friedland2008}.
The physical mechanism behind this effect is adiabatic, nonlinear
phase-locking between the system and the drive, yielding a controllable
excitation as the system's resonance frequency follows the drive frequency
as a function of time. This effect is utilized in a wide variety of
systems \cite{Andresen2010,Fajans1999}, and recently in Josephson-based
oscillators \cite{Naaman2008,Murch2011}. In a \emph{quantum} anharmonic
oscillator, the expected time evolution under a similar drive is sequential
excitation of single energy levels of the system, or {}``quantum
ladder climbing'' \cite{Lin1998}. In practice, for a given anharmonicity
the drive itself introduces some mixing between the energy levels
due to power broadening and finite bandwidth, which may wash out ladder
climbing and lead to a classical behavior in a quantum system \cite{Marcus2004,Barth2011}.
In this letter, we measure the \emph{dynamics} in these two distinct
regimes in the same system by varying the drive parameters and the
system's anharmonicity.

\begin{figure}
\begin{centering}
\includegraphics[bb=14bp 18bp 272bp 185bp,scale=0.9]{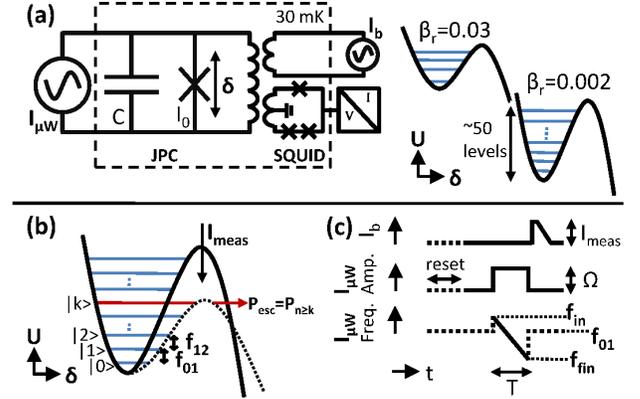}
\par\end{centering}

\centering{}\caption{\label{fig:fig1}Operation and measurement of the Josephson phase
circuit. (a) Schematics of the circuit and the potential energy at
different operating biases. The potential shape and anharmonicity
$\beta_{r}$ are set by the current source $I_{b}$ and the state
inside the well is controlled by the microwave drive $I_{\mathrm{\mu w}}$.
(b) State measurement. A short pulse $I_{\mathrm{meas}}$ is applied
in the flux bias to selectively tunnel excited levels $n>k$. The
average phase $\delta$ is then measured with an on-chip SQUID to
detect tunneling events. To determine the occupation probabilities
of all the $N$ levels, this process is repeated with a series of
different $I_{\mathrm{meas}}$ amplitudes \cite{Supp}. (c) Time sequence
of the chirp experiment. The drive amplitude $\Omega$ is expressed
in units of the Rabi frequency, measured on the first transition.}
\end{figure}

Our system, the Josephson phase circuit (JPC, see Fig. \ref{fig:fig1}a),
is a superconducting oscillator with a nonlinear inductor formed by
a Josephson junction. It can be described energetically by a
double-well potential that depends on the phase difference $\delta$
across the junction. We tune the potential by means of an external
magnetic flux bias \cite{Martinis2002} to vary the anharmonicity
and measure the state. Traditionally, the circuit is operated as a
two-level system (qubit) \cite{Martinis2002,Shevchenko2008}, or a
d-level system (qudit) \cite{Neeley2009}, by localizing the phase
$\delta$ in a shallow well where there are only a few energy levels.
The quantum state of these levels is then controlled by applying nearly
resonant current pulses. Due to the finite coherence time of the system,
this generally requires the anharmonicity inside the well $\beta_{r}=\left(f_{01}-f_{12}\right)/f_{01}$
(where $f_{ij}$ is the transition frequency from level $i$ to level
$j$) to be sufficiently large \cite{Lucero2008}. In this work, we
vary the anharmonicity over a large range ($0.002<\beta_{r}<0.03$)
in order to tune the system between the autoresonance and ladder climbing
regimes. The occupation probabilities are determined by measuring
the amount of tunneling out of the well due to a short pulse in the
flux bias that adiabatically reduces the potential barrier (see Fig.
\ref{fig:fig1}b); because of the exponential dependence of the tunneling
rate on the barrier height, we get a high tunneling contrast between
the states \cite{Neeley2009,Supp}. Tunneling events are detected
using an on-chip superconducting quantum interference device (SQUID)
\cite{Clarke2008} . The experiment is repeated $\thicksim10^{3}$
times to yield the occupation probability.

\begin{figure}
\centering{}\includegraphics[bb=14bp 16bp 266bp 358bp,clip,scale=0.93]{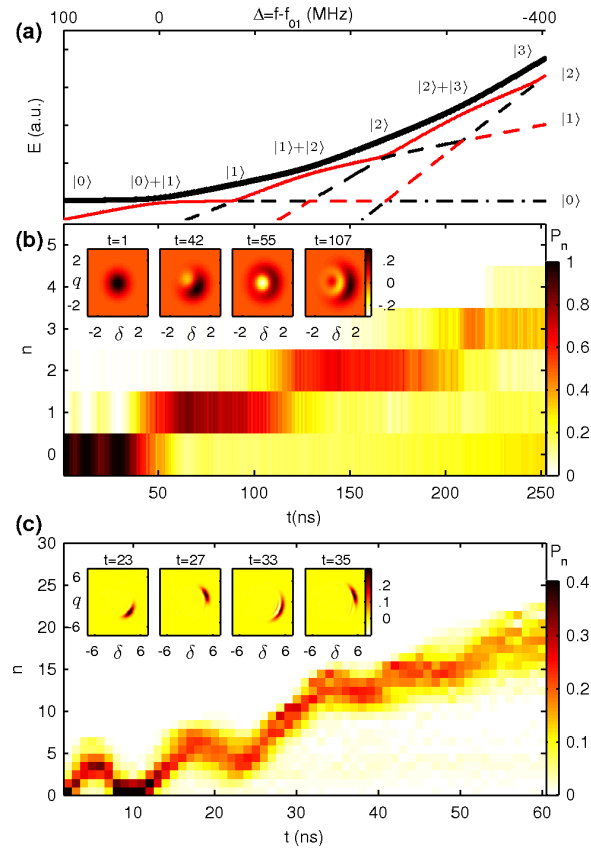}\caption{\label{fig:fig2}State dynamics during the chirp. (a) Dressed energies
of the lowest levels in the rotating frame as a function of the drive
frequency detuning $\Delta$ from the first transition $f_{01}$.
As the chirp progresses (decreasing $\Delta$), for a sufficiently
small chirp rate the state remains on the adiabatic branch (solid
black line). (b) Measured occupation probability (color-scale) as
a function of time and level number in the ladder climbing regime
($\beta_{r}=0.023$, $\alpha/2\pi=2$\,MHz/ns, $\Omega/2\pi=27$\,MHz)
and (c) autoresonance regime ($\beta_{r}=0.002$, $\alpha/2\pi=10$\,MHz/ns,
$\Omega/2\pi=190$\,MHz). The detuning scale in (a) and the time
scale in (b) are bound by the start and the end of the chirp. Insets:
simulated Wigner distribution at different times along the chirp.}
\end{figure}

The time sequence of the experiment is sketched in Fig. \ref{fig:fig1}c.
Our system has negative anharmonicity ($f_{12}<f_{01}$). Therefore,
we \emph{decrease} the drive frequency at a constant rate $\alpha=2\pi df/dt$,
starting higher than the first resonance ($f_{01}$), in accordance
with the phase locking condition. The chirp is followed by a measurement
pulse in the flux bias $I_{\mathrm{meas}}$ and the escape probability
is measured. This process is repeated for different measurement amplitudes
in order to extract the state occupation probabilities $P_{\mathrm{n}}$
\cite{Supp}. We start measuring the dynamics at a large anharmonicity
$\beta_{r}=0.023$. The time evolution is easily understood by looking
at the dressed energies of the system in the rotating frame \cite{Milburn2008}
(see Fig. \ref{fig:fig2}a). We start the chirp in the positive detuned
side ($f>f_{01}$), with the system initialized at the ground state.
As the chirp progresses (decreasing detuning), it reaches an avoided-level
crossing, associated with the first transition, at the frequency $f=f_{01}$.
If the chirp rate $\alpha$ is small relative to the splitting introduced
by the drive, an adiabatic transition \cite{Zener1932} (Landau-Zener
transition) to the 1$^{st}$ excited level occurs. As the chirp continues,
the probability of staying on the adiabatic branch (ladder climbing)
is higher than in the previous transition due to the increased energy
splitting at higher transitions ($f=f_{i,i+1}$). Figure \ref{fig:fig2}b
shows the processed data of $P_{\mathrm{n}}$ vs. time along the chirp
for the relevant states $n$. We clearly observe steps in the occupation,
corresponding to the ladder climbing effect. In phase space (see insets
of Fig. \ref{fig:fig2}b for Wigner distribution calculated from simulation),
the phase is delocalized during each step, as expected from a Fock-type
state ($\left|\psi\right\rangle =\left|n\right\rangle $). In between
the steps, there is a partial localization of the phase due to the
interference of two such states. The fidelity of each step in the
experiment (the degree of correspondence with a Fock-type state) decreases
as the state number $n$ is increased, as a result of the chirp time
being comparable to the energy decay time $(T_{1}$) of the first
excited state.

Next, we measure the evolution during a similar chirp but at a much
smaller anharmonicity - $\beta_{r}=0.002$. Lowering the anharmonicity
brings about more mixing between the levels for a given drive, and
may therefore result in the simultaneous excitation of many levels.
Figure \ref{fig:fig2}c shows the measured time evolution under these
conditions. Instead of sharp steps, we notice a broad excitation during
the chirp, consisting of up to 6 levels. On top of that, we observe
large amplitude oscillations, as expected from autoresonant wavepacket
dynamics \cite{Fajans2001}. The oscillations are also seen in phase
space simulation (see inset of Fig. \ref{fig:fig2}c) where the phase
of the localized distribution (crescent shape) oscillates during the
chirp. A detailed comparison between data and simulation, made without adjustable parameters is shown in \cite{Supp}.

\begin{figure}
\begin{centering}
\includegraphics[bb=14bp 15bp 259bp 410bp,clip,scale=0.93]{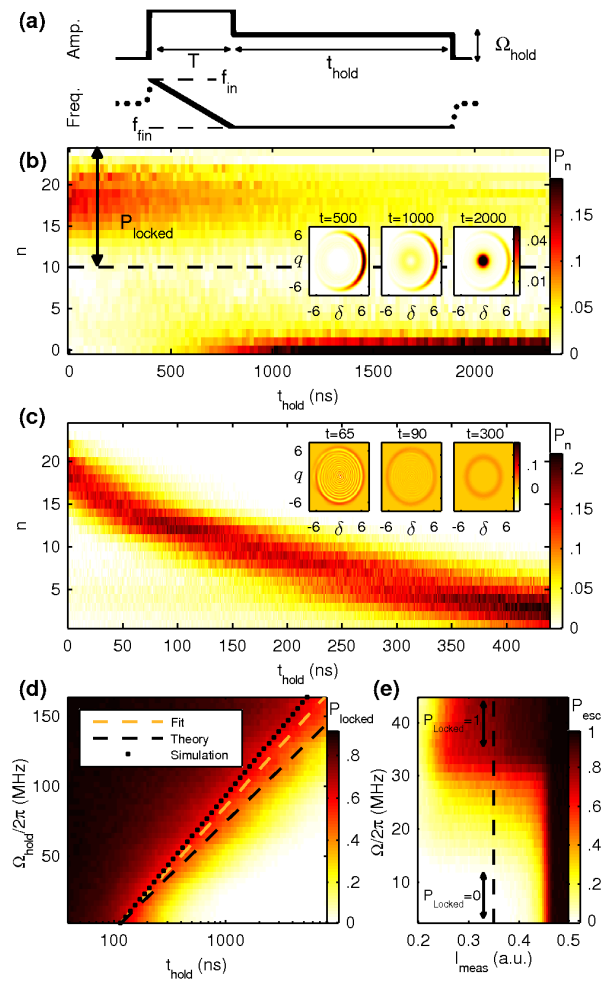}
\par\end{centering}

\centering{}\caption{\label{fig:fig3}Decay of a wavepacket. (a) Time sequence of the decay
measurement after the chirp. (b) Measured occupation probability (color-scale)
as a function of level number and time after the chirp shown in Fig.
2c, with $\Omega_{\mathrm{hold}}/2\pi=190$\,MHz and (c) $\Omega_{\mathrm{hold}}=0$.
Insets of (b) and (c) show the simulated Wigner plot at different
times along the decay. (d) Measured locking probability (color-scale)
as a function of time and amplitude of the drive after the chirp,
with contours corresponding to $P_{\mathrm{locked}}(t_{\mathrm{hold}},\Omega_{\mathrm{hold}})=0.5$,
obtained from data, theory and simulation. (e) Escape probability (color-scale) as a function of measurement
amplitude $I_{\mathrm{meas}}$ and drive amplitude $\Omega$ after
a chirp, with $\alpha/2\pi=10$\,MHz/ns and $\beta_{r}=0.0046$.
To measure the locking probability, an intermediate $I_{\mathrm{meas}}$
is used (dashed line) at the end of the chirp. }
\end{figure}

To check the stability of the generated wavepacket at small anharmonicity,
we fix the amplitude and frequency of the drive at the end of the
chirp to their final value (illustrated in Fig. \ref{fig:fig3}a,
in the case $\Omega_{\mathrm{hold}}=\Omega$, where $\Omega_{\mathrm{hold}}$
is the drive amplitude after the chirp). Figure \ref{fig:fig3}b shows
the resulting time evolution after the chirp. The phase-locked wavepacket
is centered around $n\approx18$ and is remarkably long-lived, despite
the short decay time at these highly excited levels. We define the
locking probability $P_{\mathrm{locked}}$ as the probability to be
in the phase-locked state, taken for this measurement as the integrated
probability for levels $n>10$ \cite{Supp,Barth2011}. The locking probability decays non-exponentially with
a time constant $T_{\mathrm{locked}}=1.4\,\mu$s, where $T_{\mathrm{locked}}$
is defined as the time it takes for the locking probability to decay
to half of its initial value. The results of this experiment should
be contrasted with the measurement shown in Fig. \ref{fig:fig3}c,
where $\Omega_{\mathrm{hold}}=0$. In this measurement, the energy
expectation (proportional to the average level number) decays exponentially
at roughly $T_{1}\approx300$\,ns, consistent with the expected decay
of a wavepacket in a nearly harmonic oscillator \cite{Wang2008}.
In phase space (insets of Fig. \ref{fig:fig3}c) there is a quick (5\,ns) delocalization into a pattern of circular fringes due to the non-negligible anharmonicity. The short lifetime-limited dephasing at $\left\langle n\right\rangle =18$ smears out this pattern into a ring (30\,ns) \cite{Wang2009}, shrinking at a constant rate
$\Gamma_{1}=1/T_{1}$, as expected. When $\Omega_{\mathrm{hold}}=\Omega$
(see insets of Fig. \ref{fig:fig3}b), the locked population (crescent
shape) remains localized, but slowly leaks out through the edge to
the unlocked state, which freely decays as in Fig. \ref{fig:fig3}c.

The results are explained within an effective barrier model \cite{Dykman1988,Dykman2005},
where, the drive at the end of the chirp and the system's anharmonicity form an effective potential
barrier for the population that is locked by the chirp. In this picture,
the size of the potential barrier scales as the amplitude of the drive.
We find from this theory that the resulting lifetime of the locked
population is given by $T_{\mathrm{locked}}\varpropto\exp(\eta\Omega_{\mathrm{hold}}/2\pi)$
\cite{Supp,Dykman1988}, where the parameter $\eta$ depends on the
system and drive frequencies \cite{Supp}. To check this model experimentally,
we measure the locking probability as a function of time after the chirp and of drive
amplitude. In this measurement (see Fig. \ref{fig:fig3}d) the chirp
parameters are fixed, but the drive amplitude at the end of the chirp
is varied \cite{FN1}. We find that $T_{\mathrm{locked}}$ scales
exponentially with $\Omega_{\mathrm{hold}}$, supporting the effective
barrier picture. The holding lifetime increases by nearly two orders
of magnitude to more than 10\,$\mu$s. The factor $\eta$ we extract
from this data ($\eta\approx26$\,ns), is in agreement with theoretical
prediction ($\eta\approx30$\,ns) and simulation ($\eta\approx24$\,ns)
\cite{Supp}. Note that in this experiment, the chirp is used to prepare the initial locked state only.

The locking probability is directly measured using a calibrated measurement pulse. In Fig. \ref{fig:fig3}e, as the drive amplitude is increased near the threshold ($\Omega_{th}/2\pi\approx30$\,MHz), the highly excited (phase-locked)
levels become more populated, as indicated by the increased escape
probability at smaller measurement amplitudes. To measure
the locking probability $P_{\mathrm{locked}}$, we use a
measurement amplitude that causes only the population in the upper levels to tunnel out (dashed line).

Although the state dynamics during the chirp is fundamentally different
at large and small anharmonicities, it has common features in both
regimes. In addition to the notable increase of the system's energy
at relatively small drive amplitudes, both autoresonance and ladder
climbing have a threshold in amplitude for phase-locking. While in
autoresonance the threshold amplitude $\Omega_{th}$ scales as $\alpha^{3/4}$,
in the ladder climbing regime $\Omega_{th}\varpropto\alpha^{1/2}$.
The change in scaling indicates a transition between
the two regimes \cite{Marcus2004}. To map the transition, we measure
the locking probability as a function of chirp rate, drive amplitude
and anharmonicity.

\begin{figure}
\centering{}\includegraphics[bb=14bp 16bp 262bp 245bp,scale=0.93]{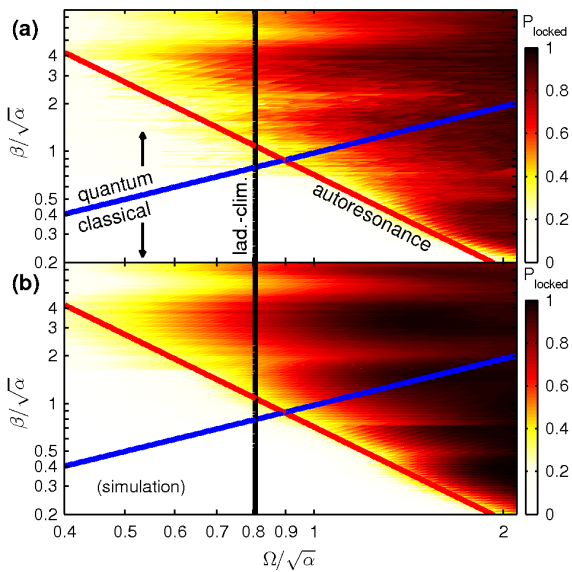}\caption{\label{fig:fig4}Transition from autoresonance to ladder-climbing.
(a) Measured locking probability
(color-scale) as a function of the dimensionless chirp parameters
$\Omega/\sqrt{\alpha}$ and $\beta/\sqrt{\alpha}$ . The red and black
lines are the theoretical thresholds for autoresonance ($\Omega_{th}^{ar}=0.82\alpha^{3/4}\beta^{-1/2}$)
and ladder climbing ($\Omega_{th}^{lc}=0.8\alpha^{1/2}$) \cite{Barth2011}.
The blue line ($\Omega=\beta$) marks
the separation between the quantum and classical regimes \cite{Marcus2004}.
(b) A simulation of the experiment shown in (a) with the same parameters,
including the effects of decay and measurement at different $\beta_{r}$
\cite{Supp}.}
\end{figure}

Following Marcus et al. \cite{Marcus2004} we plot the results (see
Fig. \ref{fig:fig4}a) in the dimensionless parameters space, $\Omega/\sqrt{\alpha}$
and $\beta/\sqrt{\alpha}$, where $\beta=2\pi\beta_{r}f_{01}$ is
the absolute anharmonicity \cite{FN2}. The measured threshold, defined
by $P_{\mathrm{locked}}(\Omega/\sqrt{\alpha},\beta/\sqrt{\alpha})=0.5$,
changes scaling (the dependence of threshold amplitude on chirp
rate) at thresholds where $\beta\approx\Omega$ (blue line). This condition is met when the broadening of the first transition (caused by the drive amplitude) is comparable to the frequency difference between neighboring transitions. This marks the transition between the classical and quantum regimes, where the energy levels are mixed or resolved \cite{Marcus2004,Barth2011}. For comparison,
the theoretical threshold lines of autoresonance and ladder climbing
are shown on the same axes in red and black respectively. Our data
converges to the theoretical scaling at the classical limit. At the quantum limit the threshold shows slow oscillations as a function of $\beta/\sqrt{\alpha}$, centered on the theoretical ladder-climbing threshold line with superimposed fast oscillations \cite{Supp}. The slow oscillations are reproduced by numerical simulation
(see Fig. \ref{fig:fig4}b) and are the result of multi-level Landau-Zener
tunneling effects \cite{Barth2011}. In the simulation, the amplitude of these oscillations
decreases at larger $\beta/\sqrt{\alpha}$ values, converging to the
theoretical ladder climbing threshold scaling \cite{Barth2011}.

In conclusion, the ability to measure the system's dynamics in different
regimes relies on the wide-range tunability of the Josephson phase
circuit. This tunability opens the possibility of measuring the full
state (state tomography) of wavepackets in more coherent devices in
the future. Using chirps, one can then generate and measure {}``cat-states''
\cite{Zurek2003}, within this macroscopic system. In the ladder climbing
regime, one can use the chirp to generate high fidelity $\left|n\right\rangle $
states in lifetime-improved devices, without the long calibration
process that is commonly required. This demonstrates the usefulness
of chirped drive in creating and manipulating quantum states in the
tunable Josephson phase circuit, with applications in rapid state
preparation and measurement.

This work was supported by ISF grant 1248/10 and BSF grant 2008438.


\begin{thebibliography}{10}
\bibitem[27]{FN1}This measurement is carried out with a different
sample. The results are similar in both samples.
\bibitem[28]{FN2}The complete data set is obtained from measurements
at four different anharmonicities ($\beta_{r}=0.026,0.022,0.004,0.002$),
in order to span a large range of $\beta/\sqrt{\alpha}$ while keeping
the chirp time $T=2\pi\left(f_{\mathrm{in}}-f_{\mathrm{fin}}\right)/\alpha$
shorter than $\sim T_{1}/2$, avoiding the effects of decay. We find
that the threshold amplitude is affected by the decay for longer chirp
times. The finite coherence time $T_{2}$ only weakly affects the
threshold in our experiment \cite{Supp}.
\end{thebibliography}
\end{document}


\title{Quantum and Classical Chirps in an Anharmonic Oscillator}

\author{Yoni Shalibo, Ya'ara Rofe, Ido Barth, Lazar Friedland, Radoslaw Bialczack,
John M. Martinis and Nadav Katz}

\maketitle
\textbf{\Large Supplementary Information}\\
{\Large \par}

\textbf{Materials and methods}. The Josephson phase circuit \cite{Martinis2002}
used in the experiment has the following design parameters: critical
current $I_{0}\approx1.5\,\mu$A, capacitance C$\approx$1.3\,pF
and inductance L$\approx$940\,pH. The qubit has a tunable frequency
$f_{01}$ in the 6-9\,GHz range \cite{FN1}. During the experiment
the device is thermally coupled to the mixing chamber of a dilution
refrigerator at 30\,mK, where thermal excitations of the qubit are
negligible. 

We use a custom built arbitrary waveform generator (AWG) having a
fast (1\,ns time resolution), 14-bit digital-to-analog converter
to produce both the chirp signal and the measurement pulse. To produce
the chirp, we modulate a high-frequency oscillator, having a frequency
$f_{\mathrm{LO}}$, using an IQ-mixer. The modulation signals, produced
by the AWG, are fed into the I and Q ports of the IQ-mixer to give
$\sqrt{I(t)^{2}+Q(t)^{2}}\cos\left(2\pi f_{\mathrm{LO}}t+\phi\right)$
at its output, where $\phi=\arctan\left(Q(t)/I(t)\right)$. To produce
a frequency shift from the high-frequency oscillator, we keep the
amplitude at the output constant while varying the phase $\phi$ linearly
in time; to produce a chirp, we use an accelerating phase: $\phi=2\pi f_{0}t-\alpha t^{2}/2$,
where $f_{0}=f_{\mathrm{in}}-f_{\mathrm{LO}}$, $f_{\mathrm{in}}$
is the initial frequency of the chirp, and $\alpha$ is the chirp
rate.

To properly measure the locking probability $P_{\mathrm{locked}}$,
it is generally desirable to have the maximal possible chirp bandwidth
$\Delta f=f_{\mathrm{in}}-f_{\mathrm{fin}}$ ($f_{\mathrm{fin}}$
being the final frequency of the chirp) in order to raise the energy
expectation of the locked population higher. This leads to a better
distinguishability between the locked and the unlocked population
at the end of the chirp and correspondingly to an increased measurement
fidelity of the locking probability, as illustrated in Fig S1. The
AWG's bandwidth limitation results in an error of up to $\sim$10\,\%
in $P_{\mathrm{locked}}$ at large anharmonicity (Fig. 1a), however
it does not affect the threshold position in Fig. 4a. We use the maximal
bandwidth (600\,MHz), varying the modulation frequency from 300\,MHz
to -300\,MHz \cite{FN2}, and setting the oscillator frequency $f_{\mathrm{LO}}$
200\,MHz lower than the qubit frequency $f_{01}$. The additional
100\,MHz of bandwidth beyond the qubit frequency is taken to reduce
the sensitivity of the threshold to initial condition \cite{Friedland2008}.

\renewcommand{\thefigure}{S\arabic{figure}}
\begin{figure}
\centering{}\includegraphics[bb=14bp 14bp 280bp 135bp]{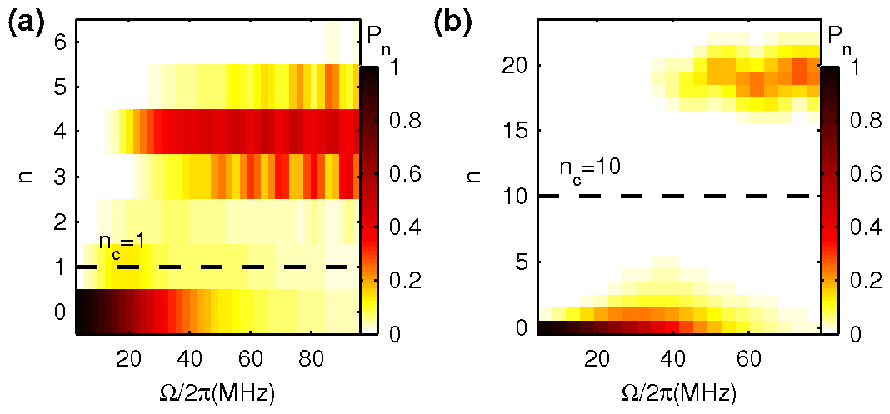}\caption{Level occupation of a bifurcated state after chirp in simulation,
as a function of amplitude. Chirp parameters: (a) $\beta_{r}=0.023$,
$\alpha/2\pi=6$\,MHz/ns and (b) $\beta_{r}=0.002$, $\alpha/2\pi=12$\,MHz/ns.
$\Delta f=600$\,MHz and $f_{01}-f_{\mathrm{fin}}=500$\,MHz in
both. The locked and unlocked populations are discriminated by a level
cutoff $n_{c}$ (dashed lines), which is experimentally realized by
a calibrated measurement pulse (see dashed line in Fig. 3e). At large
anharmonicity relative to the chirp bandwidth (a), the locked and
unlocked populations partially overlap, leading to a maximal error
of $\sim$10\,\% in $P_{\mathrm{locked}}$ for the parameters used
in the experiments.}
\end{figure}

\textbf{Data processing.} To extract the state occupation probabilities
$P_{\mathrm{n}}$, we use the escape probabilities vs. measurement
amplitude data ({}``escape curve'', $P_{\mathrm{esc}}(I_{\mathrm{meas}})$).
We first measure the single-level escape curves by preparing the system
in an $\left|n\right\rangle $ state, and then measuring the escape
probability as a function of $I_{\mathrm{meas}}$ (see Fig. S2a).
Once the single-level escape curves $P_{\mathrm{esc}}^{\mathrm{n}}(I_{\mathrm{meas}})$
are at hand, we decompose the measured escape curve of an arbitrary
state into the single-level basis $P_{\mathrm{esc}}^{\mathrm{n}}(I_{\mathrm{meas}})$
by optimizing the solution $P_{\mathrm{n}}$ to the set of $J$ equations
$P_{\mathrm{esc}}(I_{\mathrm{meas}}^{j})=\underset{n}{\sum}P_{\mathrm{n}}P_{\mathrm{esc}}^{\mathrm{n}}(I_{\mathrm{meas}}^{j})$,
where $j=1,..,J$. Generating the $\left|n\right\rangle $ state becomes
increasingly difficult at a larger $n$, due to the short lifetime
of excited states. The procedure is even more problematic when the
anharmonicity $\beta$ is small and longer pulses are required to
create the target state with reasonable fidelity. In practice, at
the small anharmonicity regime that is used in the state dynamics
measurement (see main paper), where $\beta/2\pi=18$\,MHz, it becomes
impossible to prepare the system in an $\left|n\right\rangle $ state,
even for $n>1$. Instead, we use the first excited state escape curve,
shifted by $\delta I_{\mathrm{meas}}(n)=I_{\mathrm{meas}}(0)-I_{\mathrm{meas}}(n)$
as an approximate escape curve. This approximation is supported by
WKB calculation (see below). To determine the position of the escape
curves we use the chirp data itself: for a given state, the measured
escape curve contains information about the position of the single
level escape curves. As seen in Fig. S2a, the position of these escape
curves (defined as the point where the single-level escape curve increases
to 0.5 of its maximal value) is determined from the positions of the
peaks in the derivative $\partial P_{\mathrm{esc}}(I_{\mathrm{meas}})/\partial t$.
Due to the finite width of the single level escape curves, the peak
corresponding to a certain level is visible only when the level occupation
is sufficiently large. To find $I_{\mathrm{meas}}(n)$ for all the
relevant levels, we sum the derivative over all the times along the
chirp, as illustrated in Fig. S2b. The extracted $I_{\mathrm{meas}}(n)$
values are plotted in Fig. S2c (red circles).

\renewcommand{\thefigure}{S\arabic{figure}}
\begin{figure}
\centering{}\includegraphics[bb=14bp 14bp 360bp 350bp]{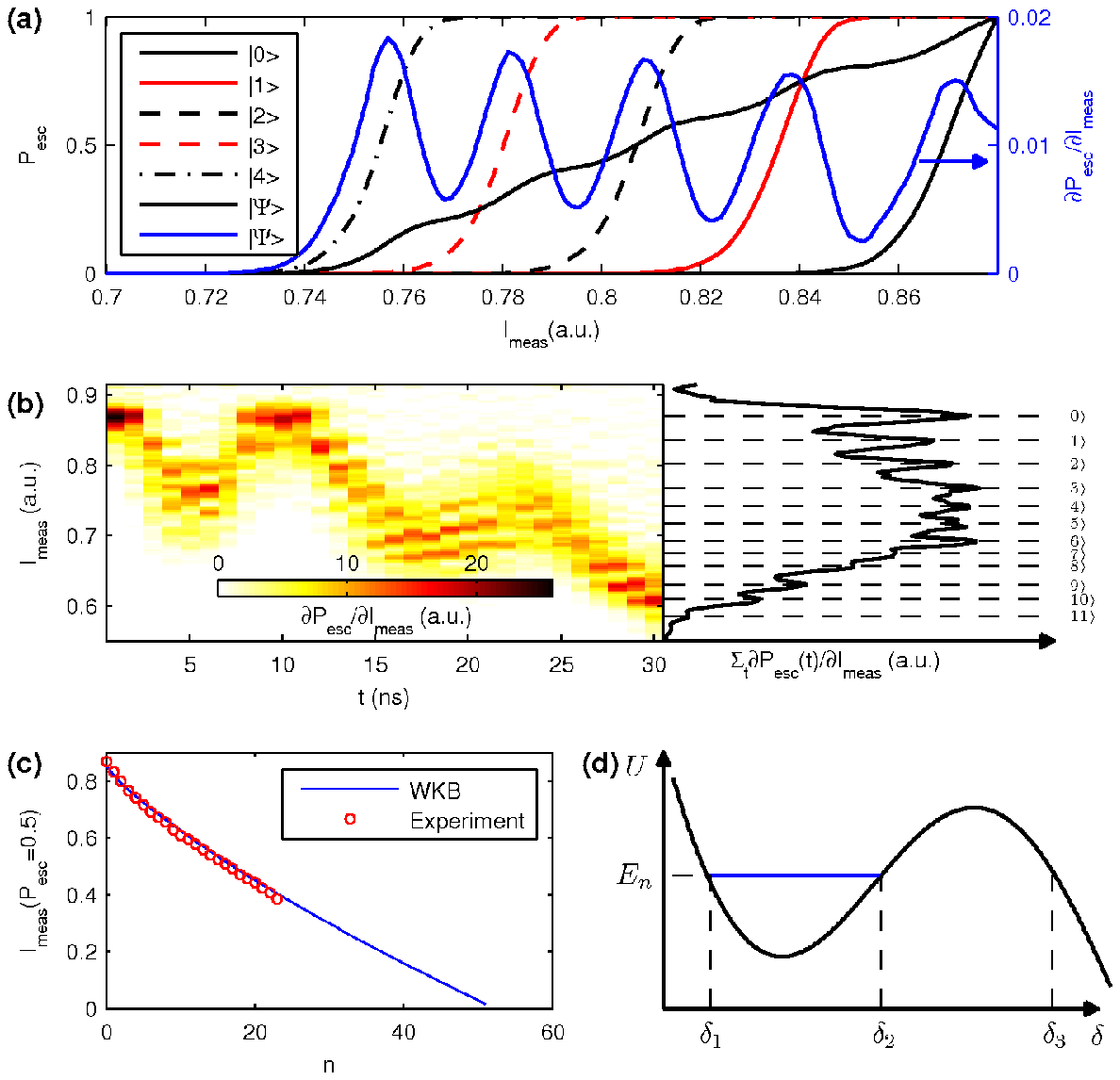}\caption{(a) Left axis: Calculated escape curves of single level states, and
of the state $\left|\Psi\right\rangle =1/\sqrt{5}\left(\left|0\right\rangle +\left|1\right\rangle +\left|2\right\rangle +\left|3\right\rangle +\left|4\right\rangle \right)$.
Right axis: derivative of the escape curve of $\left|\Psi\right\rangle $.
(b) Left panel: Derivative of the escape curve as a function measurement
amplitude and time along the chirp shown in Fig. 2c. Right panel:
Temporal sum of the data shown in the left panel, as a function of
measurement amplitude\@. (c) Experimental and calculated positions
of the escape curve. The WKB curve is calculated from the level dependent
tunneling rates, based on the calculated energies using the best fitted
circuit parameters. (d) Potential energy of the circuit used for WKB
calculation, with classical turning points.}
\end{figure}

\textbf{Simulation. }To check the validity of our estimate for the
escape curves, we calculate them numerically using the WKB approximation
of the level dependent tunneling rates \cite{Ankerhold}:

\begin{equation}
\Gamma_{\mathrm{n}}=f_{\mathrm{n}}exp(-2iS_{\mathrm{n}}),
\end{equation}
 where $S_{\mathrm{n}}=\int\limits _{\delta_{2}}^{\delta_{3}}\left|p_{\mathrm{n}}(\delta)\right|d\delta$
is the action, $\delta_{\mathrm{i}}$ are the classical turning points
defined in Fig. S2d, $p_{\mathrm{n}}(\delta)=\sqrt{2m\left(E_{\mathrm{n}}-U(\delta)\right)}$
is the momentum, $E_{\mathrm{n}}$ is the energy of the $n^{th}$
level, $U(\delta)$ is the potential energy, $f_{\mathrm{n}}$ is
the classical attempt frequency, $m=C\left(\Phi_{0}/2\pi\right)^{2}$
is the effective mass and $\Phi_{0}=h/2e$ is a flux quantum. $f_{n}$
is calculated using the classical oscillation time: $f_{\mathrm{n}}=1/\tau$
, where $\tau=\oint dt=2\int\limits _{\delta_{1}}^{\delta_{2}}\frac{d\delta}{p(\delta)/m}$
. The energies of the system are calculated by diagonalizing the system
Hamiltonian:

\begin{equation}
\hat{H}=-\frac{2e^{2}}{C}\frac{d^{2}}{d\hat{\delta}^{2}}-\frac{I_{0}\Phi_{0}}{2\pi}\cos\hat{\delta}+\frac{1}{2L}\left(\Phi_{\mathrm{ext}}-\frac{\hat{\delta}\Phi_{0}}{2\pi}\right)^{2}.\label{eq:Hamiltonian}
\end{equation}
The circuit parameters are found by best-fitting the calculated lowest
frequencies $f_{01}$ and $f_{12}$ to the measured ones and fixing
the number of levels in the well to 50 (the number of levels in the
well is obtained from extrapolating the experimental points in Fig.
S2c to $I_{\mathrm{meas}}(P_{\mathrm{esc}}=0.5)=0$) \cite{FN3}.
The single-level escape curves are then given by, $P_{\mathrm{esc}}^{\mathrm{n}}(I_{\mathrm{meas}})=1-\exp(-\Gamma_{\mathrm{n}}(I_{\mathrm{meas}})\Delta t)$,
where $\Delta t$ is the measurement pulse length. The calculated
positions of the single-level escape curves are plotted in Fig. S2c
(solid blue line).

We simulate the state dynamics of our $N$-level system under a frequency-chirped
drive by propagating its density matrix $\rho$ with the time evolution
operator $U=\exp(iH_{\mathrm{N}}\Delta t)$. The $N$-level Hamiltonian
is calculated in the rotating frame of the drive, with the rotating
wave approximation \cite{Milburn2008} applied:

\begin{equation}
H_{\mathrm{N}}=\hbar\left(\begin{array}{cccccc}
0 & \Omega/2 & 0 & 0 & \ldots & 0\\
\Omega/2 & -\Delta & \sqrt{2}\Omega/2 & 0 &  & 0\\
0 & \sqrt{2}\Omega/2 & \epsilon_{12}-2\Delta & \sqrt{3}\Omega/2 &  & 0\\
0 & 0 & \sqrt{3}\Omega/2 & \epsilon_{23}-3\Delta &  & \vdots\\
\vdots &  &  &  & \ddots & \sqrt{N}\Omega\\
0 & 0 & 0 & \ldots & \sqrt{N}\Omega & \epsilon_{\mathrm{N}-1,\mathrm{N}}-(N-1)\Delta
\end{array}\right),
\end{equation}
where $\epsilon_{\mathrm{n},\mathrm{n}+1}=2\pi\left(f_{\mathrm{n}-1,\mathrm{n}}-f_{\mathrm{n},\mathrm{n}+1}\right)$
is the anharmonicity at the $n^{th}$ level and $\Delta=\Delta(t)=2\pi\left(f(t)-f_{01}\right)$
is the frequency detuning of the drive and $h=2\pi\hbar$ is Planck's
constant. The Rabi amplitude $\Omega$ is taken as a real constant
during the chirp, and the detuning is a linearly decreasing function
starting at +$2\pi\cdot$100\,MHz and ending at -$2\pi\cdot$500\,MHz,
as done in the experiment. The anharmonicities $\epsilon_{\mathrm{n},\mathrm{n}+1}$
are calculated from the diagonalization of the system Hamiltonian
(Eq. \ref{eq:Hamiltonian}). The simulation neglects deviations of
matrix elements due to the drive, beyond the harmonic oscillator approximation.
Namely, for $m\neq n\pm1$, we set $\left\langle n\left|\hat{\delta}\right|m\right\rangle =0$,
and for $m=n\pm1$ we set $\left\langle n\left|\hat{\delta}\right|m\right\rangle =\sqrt{n+1},\sqrt{n}$
. We find for the first order matrix elements ($m=n\pm1$) a maximal
deviation of order $\sim10^{-2}$ at the largest anharmonicity, and
highest states. The second order matrix elements ($m=n\pm2$) have
a maximal value of order $\sim10^{-1}$, relative to the first order
term at the same $m$ value. Higher order elements are smaller than
$\sim10^{-4}$. For $m=n$, the contribution to the energies for the
range of drive amplitudes used in the experiment is small compared
with the rotating frame energies. A separate simulation taking into
account all the matrix elements, without the rotating wave approximation,
yields identical results in the simulations shown below (see Fig.
S3b,d) to within a $\sim10^{-2}$ deviation. Decoherence is taken
into account using quantum operations \cite{Nielsen2004} for amplitude
and phase damping. 

In Fig. S3 we plot the level populations as a function of time during
the chirp shown in Fig. 2, compared with the experimental data. The
simulation is calculated with no fit parameters and includes the effect
of energy and phase damping. The energies at large anharmonicity (Fig.
S3b) are estimated from spectroscopy data, while for small anharmonicity
(Fig. S3d) they are extracted from the diagonalization of the system
Hamiltonian. The experimental data and simulation agree qualitatively
in both regimes. At large anharmonicity, the lengths of the {}``steps''
are slightly different in the simulation due to the error in determining
the bare transition frequencies (obtained from high power spectroscopy,
where shifts and broadenings are significant). At small anharmonicity,
we see a smearing of the oscillations at higher states. This is mainly
due to the frequency dependent drive amplitude. In both measurements
(and simulations), we used a frequency dependent drive which decreases
along the chirp as $\sqrt{n(t)}$ (where $n(t)$ is the expected average
state number as a function of time) to compensate for the increasing
drive coupling at higher states which increases the mixing between
the levels. This, however, does not affect the locking condition which
is determined from the drive amplitude at the first transition.

\begin{figure}
\centering{}\includegraphics[bb=90bp 14bp 360bp 350bp]{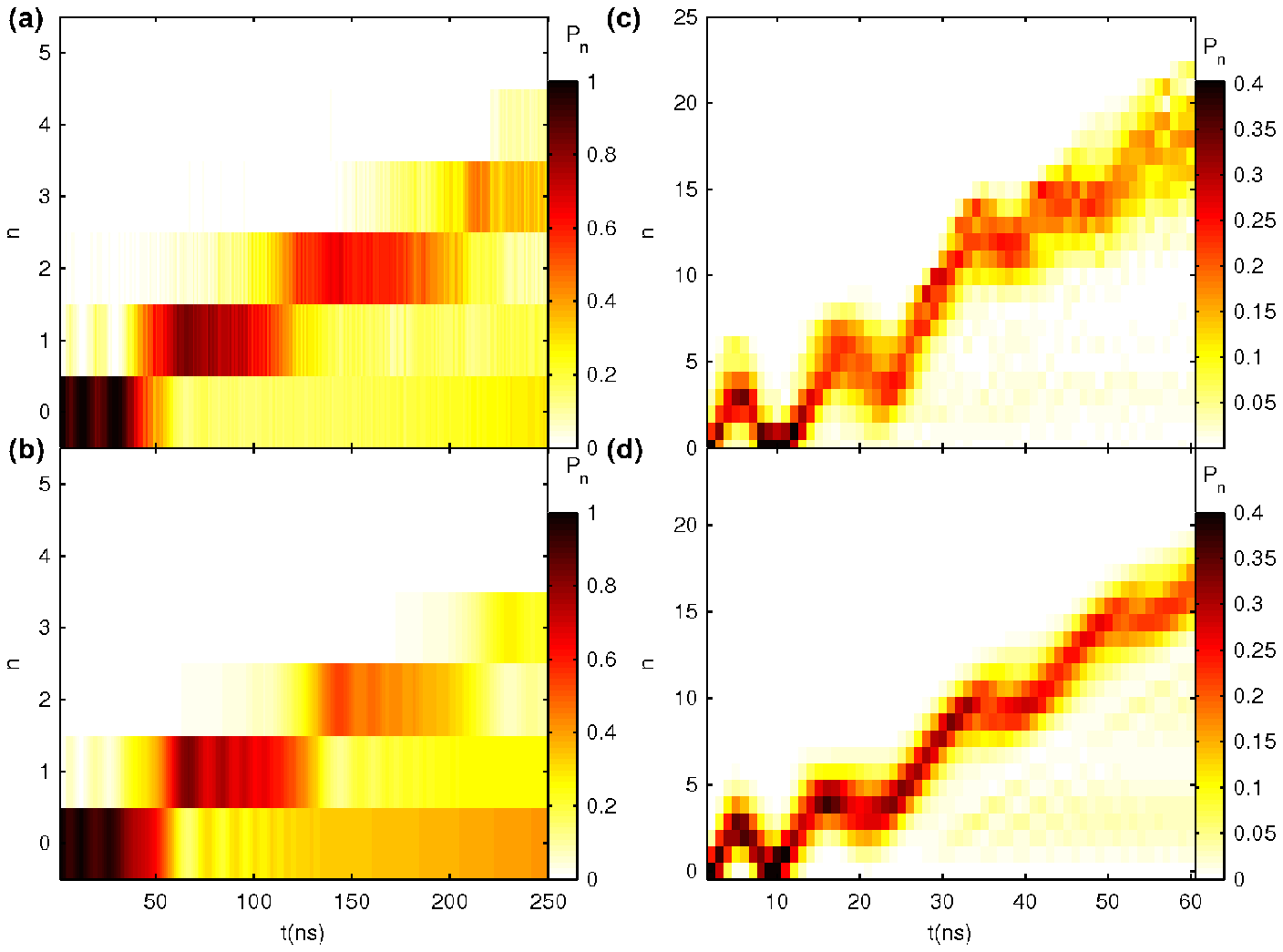}\caption{(a) Experimental data and (b) simulation of the dynamics experiment
at large anharmonicity ($\beta/2\pi$=158\,MHz) shown in Fig. 2b.
(c) Experimental data and (d) simulation of the dynamics experiment
at small anharmonicity ($\beta/2\pi$=18\,MHz) shown in Fig. 2b. }
\end{figure}

We compute the locking probability in the simulation by defining a
cutoff $n_{c}$ at intermediate levels: $P_{\mathrm{locked}}=\underset{n>n_{c}}{\sum}\rho_{\mathrm{nn}}$,
where the level population vanishes (see Fig. S1). The results of
this simulation are shown in Fig. 4b. All the parameters in the simulation
(anharmonicities, chirp rates, drive amplitudes, and decay times)
are those used/measured in the experiment. We find that the simulation
reproduces the main features of the experiment: the position of the
threshold as a function of $\beta/\sqrt{\alpha}$ and consequently
the transition between autoresonance and ladder climbing. The simuation
(as well as the experiment), displays slow averaged features with
superimposed fast oscillations. These features represent interference
between adjacent levels in the driven system. We observe that the
fast oscillations strongly depend on initial conditions (the distance
from resonance), while the averaged slow oscillations are characteristic
to the transition between the quantum and classical dynamics.

The coherence time $T_{2}$ only weakly affects the threshold in our
experiment. This claim is supported by the fact that the measured
threshold follows that of the decoherence free simulation despite
the chirp time being longer than $T_{2}$. In this simulation , we
see that at small anharmonicity, far off-diagonal elements of the
density matrix (high-order coherence terms) of a high-amplitude phase-locked
wavepacket are negligible and the phase space (Wigner) representation
of the wavepacket is similar to the one calculated for a classical
system \cite{Murch2011,Barth2011}. At large anharmonicity, we find
that only the first order coherences of the state ($\rho_{i,i+1}$
terms of the density matrix) are non-zero and they are significantly
populated for short times (compared to the relevant dephasing time),
during transitions between neighboring levels.

\textbf{Theory of held drive.} The locking time $T_{\mathrm{locked}}=W^{-1}$
(where $W$ is the decay rate from the locked state of the nonlinear
resonance) is calculated by Dykman et al. in the framework of quantum
activation \cite{Dykman1988}. It is shown that in the case of weak
damping and at low temperatures ($k_{\mathrm{B}}T\ll hf_{01}$), the
locking time is given by:

\begin{equation}
T_{\mathrm{locked}}=c\exp(\eta\Omega/2\pi),\label{eq:holding}
\end{equation}
where, $\eta\approx4/\sqrt{f_{01}\left|f_{\mathrm{fin}}-f_{01}\right|\beta_{r}}$,
and $c$ is a constant on the order of $T_{1}$. This result is valid
for intermediate drive amplitudes:

\begin{equation}
\frac{1}{2\pi T_{1}}\sqrt{\frac{4\left|f_{\mathrm{fin}}-f_{01}\right|}{\beta_{r}f_{01}}}\ll\Omega/2\pi\ll\left|f_{\mathrm{fin}}-f_{01}\right|\sqrt{\frac{4\left|f_{\mathrm{fin}}-f_{01}\right|}{\beta_{r}f_{01}}},
\end{equation}
as is the case in our experiment, where these conditions translate
to 4\,MHz$\ll\Omega/2\pi\ll$3.7\,GHz.

In this theory, the dynamics are considered to be classical while
the noise is quantum, and is associated with zero-point fluctuations.
Moreover, the expression for the locking time coincides with the classical
formula for the escape time \cite{Dykman1979}, when the classical
temperature in \cite{Dykman1979} is replaced by an effective temperature,
$T_{\text{eff}}=(hf_{01}/2k_{\mathrm{B}})\coth(hf_{01}/2k_{\mathrm{B}}T)$.
A more intuitive, but equivalent theory for the locking time is given
by Dykman et al. \cite{Dykman2005} where the escape time from an
effective potential well associated with the phase-locked state is
calculated. The potential barrier in this case scales as the drive
amplitude. 

We find good agreement between the simulation of this experiment at
several anharmonicities and the scaling predicted by Eq. \ref{eq:holding}.
The theoretical prediction of the factor $\eta$, calculated using
the experimental parameters (see black dashed line in Fig. 3d) is
within 15\,\% from that obtained in the simulation with the same
parameters.